\documentclass[letters,usenatbib]{mnras}

\usepackage[T1]{fontenc}

\DeclareRobustCommand{\VAN}[3]{#2}
\let\VANthebibliography\thebibliography
\def\thebibliography{\DeclareRobustCommand{\VAN}[3]{##3}\VANthebibliography}

\usepackage{graphicx}	
\usepackage{amsmath}	
\usepackage{amssymb}	
\usepackage{newtxtext,newtxmath}

\title[Helium abundance in ICMEs]{A holistic approach to understand Helium enrichment in Interplanetary coronal mass ejections: New insights  }

\author[Yogesh et al.]{
Yogesh,$^{1,2}$\thanks{E-mail: yphy22@gmail.com}
D. Chakrabarty$^1$,
and N. Srivastava$^3$
\\
$^1$Physical Research Laboratory, Navrangpura, Ahmedabad 380009, India\\
$^{2}$Indian Institute of Technology-Gandhinagar, Gandhinagar 382055, India\\
$^{3}$Udaipur Solar Observatory, Physical Research Laboratory, Udaipur -313001, India
}

\date{Accepted XXX. Received YYY; in original form ZZZ}

\pubyear{2022}

\begin{document}
\label{firstpage}
\pagerange{\pageref{firstpage}--\pageref{lastpage}}
\maketitle

\begin{abstract}
Despite helium abundance ($A_{He}=n_{H}/n_{He}$) is $\sim$ 8\% at the solar photospheric/chromospheric heights, $A_{He}$ can be found to exceed 8\% in interplanetary coronal mass ejections (ICMEs) on many occasions.  Although various factors like interplanetary shocks, chromospheric evaporation and “sludge removal” have been separately invoked in the past to address the $A_{He}$ enhancements in ICMEs, none of these processes could explain the variability of $A_{He}$ in ICMEs comprehensively. Based on extensive analysis of 275 ICME events, we show that there is a solar activity variation of ICME averaged $A_{He}$ values. The investigation also reveals that the first ionization potential effect as well as coronal temperature are not the major contributing factors for AHe enhancements in ICMEs.  Investigation on concurrent solar flares and ICME events for 63 cases reveals  that chromospheric evaporation in tandem with gravitational settling determine the $A_{He}$ enhancements and variabilities beyond 8\% in ICMEs. While chromospheric evaporation releases the helium from chromosphere into the corona, the gravitationally settled heliums are thrown out during the ICME eruptions. We show that the intensity and timing of the preceding flares from the same active region from where the CME erupts are important factors to understand the $A_{He}$ enhancements in ICMEs.
\end{abstract}




\section{Introduction}
The abundance of Helium with respect to hydrogen, expressed as $A_{He}=n_H/n_{He}$ \% in general, varies significantly in different layers of the Sun. While $A_{He}$ is $\sim$ 8\% in the photosphere, it remains 4-5\% in the corona. On the other hand, $A_{He}$ varies from 2 to 5\% in the quiet solar wind depending upon the phase of the solar activity and solar wind velocity \citep{Kasper2007, Alterman2019,Yogesh2021}. Interestingly, on many occasions, $A_{He}$ is found to increase significantly and exceed the photospheric abundance of $\sim$8\% \citep{Grevesse1998, Asplund2009} in the interplanetary coronal mass ejection (ICME) structures. This suggests other processes are operational for the elevated $A_{He}$ abundances in the ICME structures. The nature of these processes is poorly understood till date \citep{Manchester2017}. While elevated $A_{He}$ is observed in many ICMEs, the absence of $A_{He}$ enhancements is also noted in some ICMEs. This unresolved dichotomy is also intricately connected with the question of efficacy of $A_{He}$ as one of the ICME indicators \citep{Hirshberg1972,Borrini1982,Zurbuchen2006} in the heliosphere. 

In the past, enhanced $A_{He}$ at 1 AU has been found to be associated with interplanetary (IP) shocks \citep[e.g.,][]{Borrini1982}. However, $A_{He}$ enhancement without preceding IP shock has also been found in many cases \citep[e.g.,][]{Fenimore1980}. Based on the high ionization temperature associated with the $A_{He}$ enhancement events, \cite{Borrini1982} suggested that enhanced $A_{He}$ in the solar wind at 1 AU indicates the arrival of plasma ejecta from the solar eruptive events. As the ejecta are thrown from the lower corona, these bring out additional loads of gravitationally settled Helium \citep{Hirshberg1970}.
Therefore, the 'sludge effect' proposed by \cite{Neugebauer1997} is primarily an extension of this gravitational settling argument. Since gravitational settling is always present, it is not clear how the sludge effect can selectively enhance $A_{He}$ in certain CMEs only. On the other hand, \cite{Fu2020}  suggested important role of chromospheric evaporation for the enhancement in $A_{He}$ in CMEs. However, as the photospheric abundance of $A_{He}$ is $\sim$8\% \citep{Grevesse1998}, it is not clear how this process can enhance $A_{He}$ beyond 8\%. Therefore, despite the important roles of the above processes being qualitatively acknowledged, the relative roles of these processes governing the variability of $A_{He}$ enhancements, particularly beyond 8\%, remain elusive till date. In this work, we evaluate all these processes in totality and show that primarily chromospheric evaporation along with gravitational settling control the variabilities of $A_{He}$ in ICMEs. 

\section{Data and ICME selection}
The measurements from the Solar Wind Ion Composition Spectrometer (SWICS) \citep{Gloeckler1998} instrument onboard ACE (Advanced Composition Explorer) satellite is used in the present study. We have used the two-hourly data of different elemental compositions and charge states here.  This data set contains data from 04 February 1988 to 21 August 2011 for several elements and their charge states. On 22 August 2011, ACE/ SWICS entered into a different state due to hardware degradation caused by radiation-induced defects. After this incident, another approach \citep{Shearer2014} was used to get the compositional data. A few compositional observations were possible using this indirect and modified approach. Details on the data prior to and after the change in approach can be found at \url{http://www.srl.caltech.edu/ACE/ASC/level2/index.html}. The one hourly helium and hydrogen ratios are also used from the OMNI dataset \citep{King2005}. 

 	In this work, the ICME catalog (\url{http://www.srl.caltech.edu/ACE/ASC/DATA/level3/icmetable2.html}) compiled by \cite{Richardson2004,Richardson2010} is used to select the ICME events and their arrival at the L1 point.  The details regarding the selection criteria can be found in \cite{Richardson2004}.  The ICME list is classified into three different categories- 1. ICMEs with full magnetic cloud (MC, 86 events) characteristics, 2. ICMEs showing the magnetic rotation but lacking a few properties like magnetic field enhancements (referred to as partial MC, 92 events), and 3. ICME without most of the MC characteristics (termed as ejecta, 97 events). We have considered only those events for which the composition data are available for more than 6 hours. Relevant details like the start and end times of the duration of passage of the ICMEs through the L1 point are given in Richardson and Cane's catalog.  For a selected ICME event, the average $A_{He}$ between the start and end times is considered. The SOHO/LASCO CME catalog \citep[and references therein]{Gopalswamy2009} is used to find the details on the flares associated with these CMEs and the NOAA active region identifier from where the flares erupt. The SOHO/LASCO Halo CME catalog details can be found at \url{https://cdaw.gsfc.nasa.gov/CME_list/halo/halo.html}. 

\section{Results and Discussions }
\subsection{Solar activity variation of $A_{He}$ in ICMEs}
\begin{figure}
\begin{center}

\includegraphics[scale=0.55]{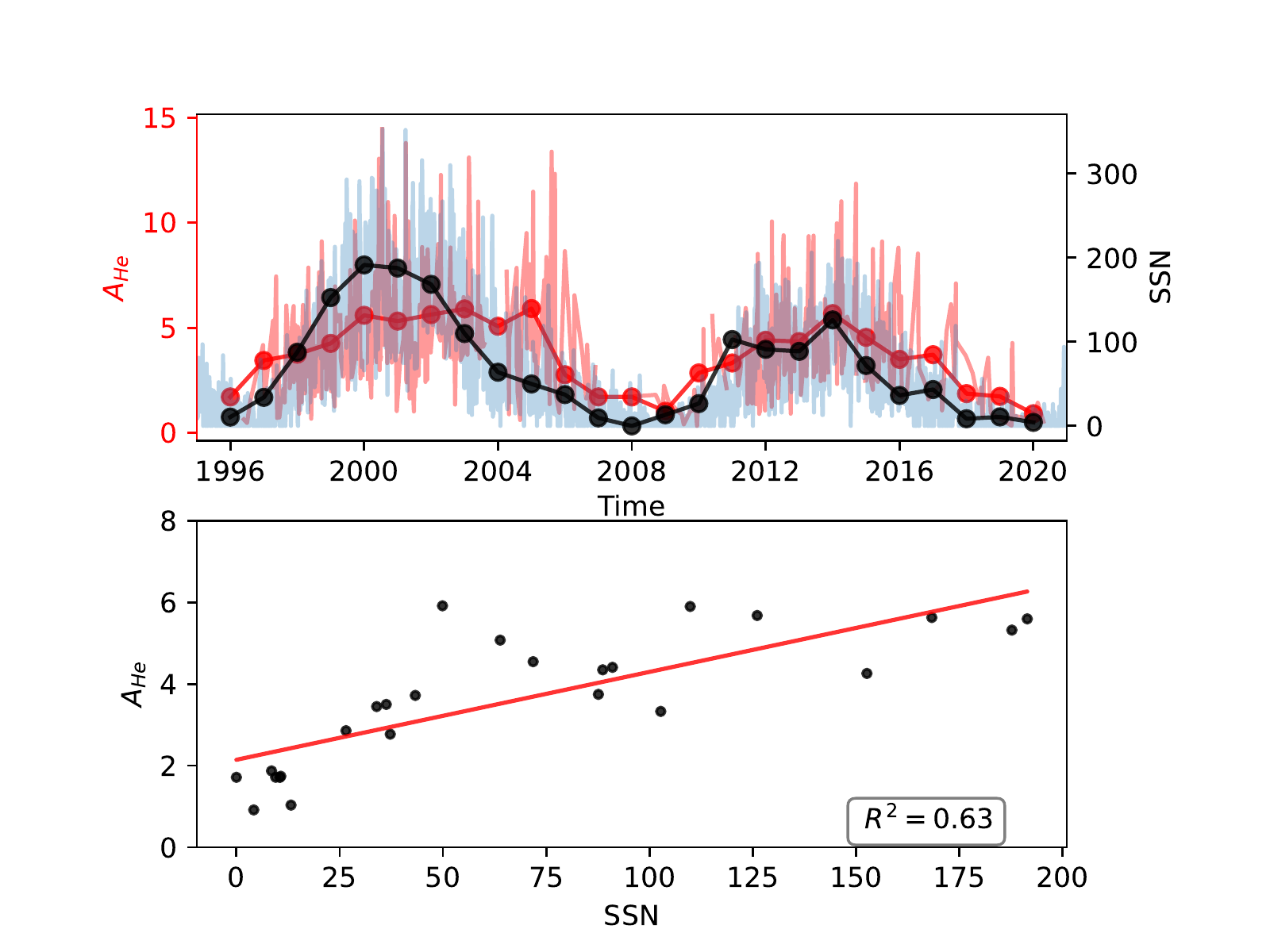}
\caption{\textbf{Solar activity variation of ICME averaged $A_{He}$ -} (a) The red and black colored lines in the background show the individual ICME averaged $A_{He}$ and sunspot number (SSN). The filled red and black circles joined by lines are the yearly averaged $A_{He}$ and SSN for the ICME duration only. (b) The correlation between the annual averaged $A_{He}$ and SSN for the ICME duration is shown.     \label{fig:1}}

\end{center}
\end{figure}

Since $A_{He}$ in solar wind varies with solar activity level \citep{Kasper2007, Alterman2019,Yogesh2021}, it is important to check whether the variation of $A_{He}$ in ICMEs also show solar activity variation. In order to address this aspect, solar cycle (Cycle 23 and 24) variation of $A_{He}$ during ICME events is analysed. In Figure \ref{fig:1}a (upper panel), the red and black lines depict the individual ICME averaged $A_{He}$ and sunspot number (SSN). The lines joining red and black dots are the yearly averaged SSN and $A_{He}$ averaged over the ICME durations for a given year. Interestingly, the yearly ICME averaged $A_{He}$ (varies from 1 to 6), and SSN show good correlation ($R^2$ = 0.63) as brought out in Figure \ref{fig:1}b (lower panel). This suggests that the processes controlling the $A_{He}$ in background solar wind modulate the ICME averaged $A_{He}$ values also. Another important point that emerges from Figure \ref{fig:1} is the higher value of ICME averaged $A_{He}$ in the year 2005. Similar kind of enhancements in the occurrence of flare \citep{Hudson2014} and CME eruptions \citep{Mishra2019} in 2004-2005 were reported earlier. This is indicative of the important role of solar flares in the $A_{He}$ enhancement. This aspect will be exclusively taken up in a subsequent section. 

\subsection{Relationship of $A_{He}$ with FIP elemental ratios, average charge states and charge state ratios}
It is, in general, observed \citep{Zurbuchen2016} that the low FIP elements are enhanced during the ICMEs as compared to the ambient solar wind. These authors suggested that either FIP bias is more significant during the CME eruption or different type of plasma is injected into the CME. Further, the magnetic energy is converted to thermal and kinetic energies through magnetic reconnection in the corona during the CME initiation process \citep{Forbes2006,Wimmer2006} and the flares associated with these events. The charge states and the charge state ratios get frozen in the corona and the CME eruption throws out these plasma from the corona into the solar wind \citep{Gruesbeck2011,Gruesbeck2012}. Therfore, it is worthwhile to evaluate the relaionship of $A_{He}$ with FIP elemental ratios, average charge states and charge state ratios.

We thus tested the relationship of $A_{He}$ with different FIP proxies (like Mg/O, Fe/O, Si/O, C/O, Ne/O, and He/O) for background solar wind, ejecta, partial MC and MC varieties of ICMEs.  As the FIP of Mg, Fe, Si, C, Ne, He, and O are 7.65, 7.9, 8.2, 11.3, 21.6, 24.6, and 13.1 eV, respectively, the correlation results of Mg/O, Fe/O, Si/O, C/O, Ne/O, and He/O with respect to $A_{He}$ for the four categories (ambient solar wind, ejecta, partial MC and MC) are summarized from top to bottom in Table 1. The corresponding plots are provided as supplementary Figure S1. Two types of correlation coefficients are considered. One is the linear correlation coefficient (CC - parameterized by the Coefficient of determination, $R^2$) and the other is the Spearman’s Correlation coefficient (Sp. CC). It is noted that the FIP proxies are uncorrelated with $A_{He}$ in the ambient  solar wind (SW). However, with the exception of He/O and C/O, $R^2$ improves for the MC category compared to the non-MC category. It is also noted that Sp CCs are higher than $R^2$ indicating control of processes other than FIP. This is because, Sp CC is a measure of monotonic relationship and does not necessarily imply a linear relationship. 
\begin{table*}
 \caption{\textbf{The results from the detailed correlatio exercises of $A_{He}$ with FIP elemental ratios, average charge states and charge state ratios} - Linear correlation coefficient (CC - parameterized by the Coefficient of determination, $R^2$) and the and Spearman’s Correlation coefficients (Sp. CC) are calculated and tabulated for ambient SW, ejecta, partial MC as well as MC. The corresponding figures are provided as supplementary materials. The correlation of $A_{He}$ in ICME is non-existent for ambient SW and maximum for MC. Importantly, Sp. CC always exceeds CC indicating the important role of other processes in determining the abundance of $A_{He}$ in ICMEs.}
 \label{tab:example}
\begin{tabular}{|c|c|c|c|c|c|c|c|c|}
\hline 
 & \multicolumn{2}{c|}{Ambient SW (184)}& \multicolumn{2}{c|}{Ejecta (97)} & \multicolumn{2}{c|}{Partial MC (92)} & \multicolumn{2}{c|}{MC (86)} \\ 
\hline 
 & \multicolumn{7}{c|}{FIP Proxies} \\ 
\hline 
 & CC & Sp CC & CC & Sp CC & CC & Sp CC & CC & Sp CC \\ 
\hline 
Mg/O &0.06   & -0.29 &0.05  &0.27  &0.19  &0.35  &0.32&0.54  \\ 
\hline 
Fe/O &0.03  &0.12  &0.13  &0.40  &0.23 &0.35  &0.33  &0.56  \\ 
\hline 
Si/O & 0.00  &-0.01  &0.06  &0.32  &0.15  & 0.33 &0.32  &0.56  \\ 
\hline 
C/O &0.08  &-0.28  &0.26  &-0.57  &0.12  &-0.38  &0.07  &0.26  \\ 
\hline 
Ne/O & 0.03 & -0.22 & 0.20 &0.52  & 0.31 &0.49  &0.42  &0.58  \\ 
\hline 
He/O &0.12  &0.36  &0.02  &0.13  & 0.09 & 0.30 & 0.15 &0.39 \\ 
\hline 
 & \multicolumn{7}{c|}{Average charge States} \\ 
\hline 
$Q_C$ & 0.01 & 0.08 & 0.02 & -0.15 & 0.02 & 0.08 & 0.00 & 0.00 \\ 
\hline 
$Q_O$ & 0.07 & 0.20 & 0.19 & 0.51 & 0.34 & 0.56 & 0.32 & 0.57 \\ 
\hline 
$Q_{Mg}$ & 0.11 & 0.33 & 0.28 & 0.63 & 0.42 & 0.65 & 0.35 & 0.57 \\ 
\hline 
$Q_{Si}$ & 0.05 & 0.21 & 0.29 & 0.58 & 0.44 & 0.66 & 0.42 & 0.65 \\ 
\hline 
$Q_{Fe}$ & 0.05 & 0.22 & 0.26 & 0.56 & 0.44 & 0.62 & 0.36 & 0.39 \\ 
\hline 
 & \multicolumn{7}{c|}{Charge state Ratios} \\ 
\hline 
$C^{+6}/C^{+4}$ & 0.01 & 0.07 & 0.00 & 0.00 & 0.05 & 0.25 & 0.03 & 0.17 \\ 
\hline 
$C^{+6}/C^{+5}$ & 0.01 & 0.06 & 0.01 & -0.08 & 0.04 & 0.15 & 0.02 & 0.12 \\ 
\hline 
$O^{+7}/O^{+6}$ & 0.06 & 0.19& 0.21 & 0.57 & 0.31 & 0.60 & 0.31 & 0.63 \\ 
\hline 
\end{tabular} 
\end{table*}

In the intermediate section of Table 1, we have tabulated the linear correlation coefficients (CC parameterized as $R^2$), and Sp. CC between the average charge state of C, O, Mg, Si, Fe and $A_{He}$, respectively. The corresponding plots are provided as supplementary Figure S2. A few important points can be inferred in this case. First,  $R^2$ is significantly less for the ambient solar wind. Second,  the $R^2$ values are higher for $Q_{Si}$, $Q_{Fe}$, $Q_{Mg}$, and $Q_O$ but nearly zero for $Q_C$. Third, $R^2$ values for ejecta are less than their MC and partial MC counterparts. The increased $R^2$ for ICMEs suggests modification in $A_{He}$ by coronal temperatures. However, as Sp CCs are more than CC values, factors other than coronal temperature for the $A_{He}$ enhancements in ICMEs are important. 
In order to further explore the role of coronal temperature for the $A_{He}$ enhacnements during ICME, CC and Sp CC between the charge state ratios of C, O and $A_{He}$ are explored. This is also captured in Table 1 and the plots are provided as supplementary Figure S3. Similar to previous cases, we again find negligible correlations exist between the charge state ratios and $A_{He}$ for the ambient solar wind. In addition, the number density of carbon charge states ($C^{+6}/C^{+4}$ and $C^{+6}/C^{+5}$) also do not show a significant correlation with $A_{He}$. However, the Oxygen charge states, $O^{+7}/O^{+6}$, show much higher correlations with $A_{He}$, particularly for MC. Similar to what has been noted earlier, the Sp CCs also show higher correlations for ejecta, partial MC, and MC in the case of $O^{+7}/O^{+6}$. This indicates contributions from factors other than coronal temperature for the $A_{He}$ enhancements in ICMEs. 
One way to verify these results is to perform tests on similar proxies and check if the correlation coefficients increase significantly. Following this line of thinking, We found very high correlation coefficient ($R^2$ =0.67 ) between Mg/O and Fe/O as well as Si/O and Fe/O ($R^2$ =0.80) in the case of MC. In addition, the CC values are also found to be closer to Sp CC as one considers ICMEs. These results are also provided as supplementary figure S4. This exercise suggests that if the underlying process is identical (in this case, FIP effect), one can expect significantly higher correlations. Further, it is possible that CC is higher for MC (Table 1) because the MCs offer well-defined flux ropes to be intercepted by the in-situ S/C while partial MC or ejecta may be a consequence of flank encounters with the S/C and thus all the properies of the ICMEs (MC) are not captured efficiently leading to relatively poor correlation.

Based on these arguments, we infer that the processes like coronal temperatures that determine the average charge states and charge state ratios or the FIP effect in the chromosphere may contribute to the processes that determine $A_{He}$ in ICMEs up to a certain degree. The enhanced Sp CC (compared to the linear CC) strongly suggests the presence of non-linear contribution from other processes. 

\subsection{Chromospheric evaporation and Sludge effect }

\begin{figure*}
\begin{center}
\includegraphics[scale=0.55]{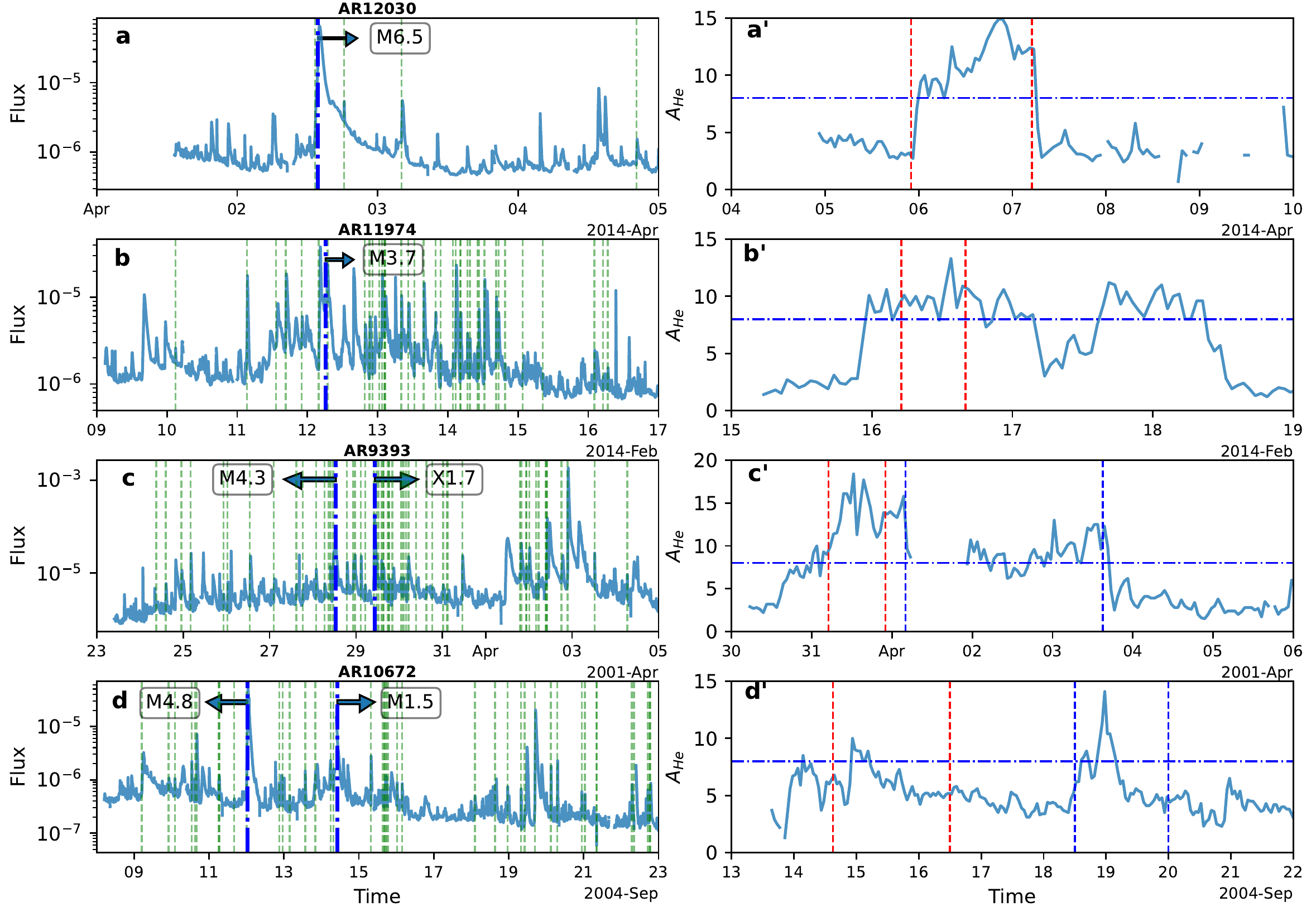}
\caption{\textbf{Solar flares and $A_{He}$ enhancements in ICMEs } – (a-d) GOES X-ray flux variation in sky blue lines for four representative cases (02 April, 2014, 12 February, 2014, 28 March, 2001 – 29 March, 2001 and 12 September, 2004 - 14 September, 2004). The dark blue vertical lines indicate the CME eruption time from the NOAA active regions from where the flares erupted. The green dashed vertical lines are the time of occurrence of flares from the same NOAA active region. The class of the flare just preceding the CME eruption is mentioned. (a’-d’) $A_{He}$ variation for the associated ICMEs at the L1 point. The vertical dashed red and blue lines mark the start and end times of the passage of ICME at the L1 point. The horizontal blue dashed lines are the $A_{He}$ = 8 level. Note, $A_{He}$ > 8\% are considered enhancements here.     \label{fig:2}}

\end{center}
\end{figure*}
Recently, it has been suggested \cite{Fu2020} that the chromospheric evaporation associated with a flare can alter the $A_{He}$ values. Interestingy, the thermal energy release during a flare can influence the charge states as well as the frozen-in signatures. Therefore, at this point, we evaluate the relationship between the occurrence of flares and observed $A_{He}$ enhancements in ICMEs. The number of flares associated with each active region are considered along with the strength of the flares during the course of the CME development. The information on the number of CMEs and the occurrence of  collocated flares for each CME are available for 63 cases. Out of these 63 cases, 17 cases are associated with X-class flares, 29 with M-class, 16 with C-class, and 1 with B-class flare. We find that 88\% (15/17) of the ICMEs associated with X-class flares show $A_{He}$ enhancement at the L1 point for at least an hour. The corresponding numbers for ICMEs with M and C-class flares are   $\sim$76\% and 50\%, respectively. Also, the MCs have highest probability (21/27 $\sim$ 78\%) of $A_{He}$ enhancement as compared to partial MC (16/22 $\sim$ 73\%) and ejecta (9/14 $\sim$ 64\%) for the 63 ICMEs. In totality, 46 out of 63 ICME events ($\sim$73\%) show $A_{He}$ > 8\% for at least one hour in presence of flares and this number increases to 59 ($\sim$94\%) if an additional 12 hours is considered beyond the ICME start and end time, respectively. As the ICME start and end times at the L1 point can differ significantly on occasions if one goes by compositional boundaries or magnetic field boundaries \citep[][also see Richardson and Cane Catalogue]{Gopalswamy2013}, consideration of additional 12 hours take care of the uncertainties involved in identifying the passage of ICMEs at the L1 point.
Since occurrence of $A_{He}$ enhancement is more likely with the strength of the flare, we infer that stronger flares lead to stronger chromospheric evaporation contributing to the higher $A_{He}$ enhancements. These statistical results are also provided as a supplementary table S5. However, chromospheric evaporation can enhance $A_{He}$ only upto the chromospheric/photospheric abundance limit (8\%). Therefore, $A_{He}$ values greater than 8\% in ICMEs at the L1 point on many occasions cannot be explained by chromospheric evaporation alone. The work by \citep{Geiss1970} suggests that $A_{He}$ accumulates in the chromosphere and/or in the lower corona by inefficient Coulomb drag exerted by protons on Helium causing the bulging of Helium in the chromosphere and/or lower corona. More importantly, helium being heavier than hydrogen, gravitational settling \citep{Hirshberg1970,Laming2019} contributes significantly to the piling up of Helium at lower coronal heights.  In fact, the large lags (of the order of 100 days) observed by \cite{Yogesh2021} strongly suggests the dominant role played by gravitational settling at lower heights. Solar wind brings out this excess Helium through CMEs mediated by process akin to “Sludge removal” \citep{Neugebauer1997} or “cleaning out” \citep{Wimmer2006}. We here suggest that chromospheric evaporation along with the 'cleaning out' of the gravitationally settled Helium enriched sludge (or, to some extent settled by inefficient Coulomb drag) CMEs can contribute to the higher values of $A_{He}$ greater than 8\%. However, it is obvious that this may not happen for all CMEs.
	To understand when the ICME events with $A_{He}$ > 8\% is a possibility, we divide the CME events into three classes and build up a statistical picture. This is captured in the supplementary Table S5. Under class-1, we consider the CMEs (20 events) with a nearly concurrent flare event prior to the CME and no previous flare activities from the same active region (except the concurrent one) for 12 hours prior to the CME. Under class-2, we consider CMEs (20 events) with multiple flares (without additional CMEs) prior to the CME from the vicinity of the same active region. The class-3 is for the multiple CMEs (23 events-11 active regions, on an average of two or more CMEs per active region) erupting from a single active region. One typical example of classes 1 and 2 is shown in the first (Figure \ref{fig:2}a-a’) and second (Figure \ref{fig:2}b-b’) row of Figure \ref{fig:2}. The third (Figure \ref{fig:2}c-c’) and fourth (Figure \ref{fig:2}d-d’) rows are examples of class 3. The left column (Figure \ref{fig:2}a-d) of Figure \ref{fig:2} shows the GOES X-ray flux variation in sky blue lines for four representative cases (02 April, 2014, 12 February, 2014, 28 March, 2001 – 29 March, 2001 and 12 September, 2004 - 14 September, 2004). The green vertical dashed lines mark the flares that originated from the same active region where the CME originated. The dark blue vertical dashed lines mark the eruption times of the CMEs. The class (mentioned in rectangular boxes) of the flares that erupted just before the CME eruption are also marked in Figures \ref{fig:2}a-d. The right column of Figure \ref{fig:2} (Figure \ref{fig:2}a’-d’) shows the variation of $A_{He}$ in the ICMEs as measured from the L1 point for the four cases shown in the left column. In \ref{fig:2}(a’-d’), the vertical dashed red and blue lines are the start and end times of the passage of ICME at the L1 point. The horizontal blue dashed lines mark the $A_{He}$ = 8 level. $A_{He}$ > 8\% are considered enhancements. We observe the highest percentage (80\%) of $A_{He}$ enhancement events falling under class-2. On the contrary, the class-1 events show the lowest percentage (65\%) of $A_{He}$ enhancement events. Class-3 shows $A_{He}$ enhancements in 74\% of the cases. Therefore, these analyses reveal that CMEs with the near simultaneous occurrence of multiple flares from the same active region (class 2) predominantly have $A_{He}$ > 8\% as compared to CMEs with a single flare from the nearby location (class 1). This suggests that the CMEs with higher Helium abundance at L1 point carry more Helium rich plasma from the lower coronal region released by chromospheric evaporation processes occurring during multiple flares. 
	
	To understand the effect of gravitational settling, the events \ref{fig:2}c and \ref{fig:2}d are chosen. These two events are selected based on the time difference between the two CME eruptions. The first event (Figure \ref{fig:2}c) has the time difference of ~21.5 hours (less than 1 day) between the two CMEs whereas the second event has a time difference of ~57.5 hrs (more than two days). As the gravitational settling time for Helium is $\sim$ 1.5 days \citep{Laming2019}, if the second CME erupts before the helium gets gravitationally settled, the second CME can be expected to have lesser Helium abundance than the first one. We speculate that this must have happened for the case shown in Figure \ref{fig:2}c’. On the contrary, if the second CME erupts sufficiently later than the first one (as in Figure \ref{fig:2}d-d’), the helium abundance in the second CME can be more (in this case) or less depending upon the accumulation of Helium. We got only two cases in our filtered database with a time difference between CMEs less than 1.5 days. More such cases in future will strengthen our argument. We note here that the timescales for the gravitational settling is more than that of chromospheric evaporation \citep[$\sim$ less than an hour,][]{Zurbuchen2016} and less than that of FIP bias \citep[a few days,][]{Zurbuchen2016}. Therefore, if an intense flare (and the associated CME) occurs at an opportune time when sufficient Helium has settled down, it will throw out significant helium into the ICMEs through chromospheric evaporation. Therefore, these resullts strongly indicate the primary role of the combined effects of chromospheric evaporation and “sludge removal” for the enhanced $A_{He}$ abundance in CMEs. Although the evidence for the combined roles of solar flare and sludge removal in $A_{He}$ enhancement is compelling, there exists a small subset of ICMEs that do not show any $A_{He}$ enhancement whatsoever. This class of ICMEs need separate investigations.  

\section{Conclusions}
This investigation shows that although solar activity variation, FIP effect, coronal temperature contribute in certain degrees towards $A_{He}$ enhancements in ICMEs at the L1 point, it is the chromospheric evaporation during solar flares assisted by gravitational settling of Helium that determines the enhancement of $A_{He}$ in ICMEs. It is shown that while chromospheric evaporation is important in releasing the helium in CMEs, gravitationally settled helium thrown out of the corona during chromospheric evaporation process helps the $A_{He}$ levels to exceed the 8\% photospheric/chrmospheric level. It is suggested that the time constants of chromospheric evaporation and gravitational settling are important parameters to understand the $A_{He}$ enhancement events. We also find ICMEs wherein $A_{He}$ enhancements beyond 8\% are not observed at the L1 point. These events require further attention.

\section*{Acknowledgements}

We would like to thank the managing team of OMNI dataset (\url{https://cdaweb.gsfc.nasa.gov/team.html}). We thank and acknowledge the significant efforts put forward by the Principal Investigators (PIs) of all the satellites, data of which are used to generate this integrated, cross-calibrated OMNI data set. We thank the PIs of SWICS (ACE) and ACE Science Centre for providing the ACE data. We acknowledge NOAA and GOES for their open data policy. This work is supported by the Department of Space, Government of India. 

\section*{Data Availability}
The data can be obtained from \url{https://cdaweb.gsfc.nasa.gov/index.html/}. The Goes X-ray flux can be seen from \url{https://satdat.ngdc.noaa.gov/sem/goes/data/avg/}.



\bibliographystyle{mnras}
\bibliography{ref} 



\label{lastpage}

\end{document}